\documentclass[runningheads]{llncs}
\usepackage{graphicx}
\usepackage{subcaption}
\usepackage{mathtools}
\usepackage{physics}
\usepackage{float}
\usepackage{hyperref}

\begin{document}
\title{Quantum Bayesian decision-making \thanks{
This work is financed by the ERDF – European Regional Development Fund through the Operational Programme for Competitiveness and Internationalisation - COMPETE 2020 Programme and by National Funds through the Portuguese funding agency, FCT, within project \texttt{POCI-01-0145-FEDER-030947}. The first author was further supported by project \texttt{NORTE-01-0145-FEDER-000037}, funded by Norte Portugal Regional Operational Programme (NORTE 2020), under the PORTUGAL 2020 Partnership Agreement.}}
%
%
\author{Michael de Oliveira\inst{1} and
Luis Soares Barbosa\inst{1}}
\authorrunning{Michael Oliveira et al.}
%
\institute{INL - Quantum Software Engineering \& Universidade do Minho, Braga, Portugal
\email{michaeldeoliveira@live.com.pt, lsb@di.uminho.pt}}
\maketitle              
\begin{abstract}
As a compact representation of joint probability distributions over a dependence graph of
random variables, and a tool for modelling and reasoning in the presence of uncertainty, 
Bayesian networks are of great importance for artificial intelligence to combine domain knowledge, capture causal relationships, or learn from incomplete datasets. Known as a NP-hard problem in a classical setting, Bayesian inference pops up as a class of algorithms worth to explore in a quantum framework.  This paper explores such a 
research direction and improves on previous proposals by a judicious use of the utility function
in an entangled configuration. It proposes a completely quantum mechanical decision-making process with a proven computational advantage. A prototype implementation in Qiskit (a Python-based program
development kit for the IBM Q machine) is discussed as a proof-of-concept. 
\keywords{Bayesian inference \and Quantum algorithms \and Quantum decision making}
\end{abstract}

\section{Motivation}
Bayesian reasoning is widely used in machine learning and data science, as a powerful framework for probabilistic analysis, applications ranging from learning processes \cite{Neal1996} to pragmatic representations \cite{Li2018}. Broadly speaking, 
machine learning algorithms are able to learn from data, with the purpose of performing some tasks, without requiring explicit programming; in a sense outcomes are directly  built by the sampled data. However, the current rate of data creation is almost exponential \cite{Al-Jarrah20155} (going, for example,  from 3.5 million text messages per minute in 2016, to over 15 million in
2017), a fact that calls for radically new approaches and, most probably, new computational models and hardware to effectively deal with such numbers.

Can quantum computing bring some useful contribution to this state of affairs? On the one hand it is well known that building very large quantum-addressable classical memories is technologically very demanding, and will not be available soon. On the other, at least from a theoretical point of view, the question seems worth to discuss. Actually, even at its present, quite preliminar stage of development, quantum computing allows for a variety of speed ups with respect to classical algorithmic counterparts in e.g.   information storage \cite{Giovannetti2008}, pattern recognition \cite{Biamonte2016}, and  matrix inversion, the latter being a basic ingredient of several machine learning algorithms \cite{Harrow2009}. As a matter of fact, quantum algorithms, as the ones discussed in this paper, suggest radically different ways to approach old problems and 
 to explore complexity boundaries. For example, to know whether for a concrete problem, as the size of the input parameter grows, one may asymptotically go faster with the use of a quantum memory than with purely classical states, is a question underlying  many interesting problems  from big-data to optimisation, or molecular synthesis.
 
 The synergies between research lines in quantum technologies and Bayesian inference, in particular, seem promising. In one direction, a quantum processor can be expressed and studied as a Bayesian Network \cite{Sakkaris2016}. In the reverse one, quantum  mechanics can describe naturally probabilistic systems in physical terms. Reference \cite{Man09}, for example, describes very promising improvements on the implementation of approximate Bayesian inference routines resorting to physical stochastic
logic gates building up hardware implementations of sampling algorithms. Quantum processors are, in a sense, part of such a family.
 
 This paper is a step in this direction. Our starting point is a quantum version of a Bayesian inference algorithm introduced by Low \emph{et al} \cite{Low2014} based on a square-root quantum speedup to rejection sampling on a Bayesian network, which avoids the use of an oracle. Note that an oracle-based version appeared previously in \cite{OzolsRR12}. This approach, revisited in section~\ref{sc:bayes}, was implemented by us on Qiskit --- the IBM open-source platform for quantum computing. Our main contribution, presented in section~\ref{sc:dm}, extends Low \emph{et al} algorithm to a decision-making setting: this incorporates an utility function which is applied before any observation of the quantum state which encodes the Bayesian network. The computational effort for the proposed solution and a simpler quantum solution are determined in Section \ref{cplx}. A proof-of-concept  implementation  Qiskit  is  discussed  in  section \ref{poc}.  Finally,  section  \ref{conc} concludes and points out a number of issues for future work. A background section — section 2 — recalls the Bayesian inference problem and provides a brief overview of the basic intuitions underlying decision making.

\section{Background}\label{sc:bk}

\subsection{Bayesian inference}

Bayesian inference is used to update the posterior probability distribution of some query variables given the value of the observed variables, also known as evidence variables \cite{artificial}. The conditional probability is given by
   	\begin{equation} \label{eq:Bayestheorem}
	P(A|B)=\frac{P(A,B)}{P(B)}
	\end{equation}

\begin{figure}

\begin{subfigure}{0.3\textwidth}
\centering
\includegraphics[scale=0.15]{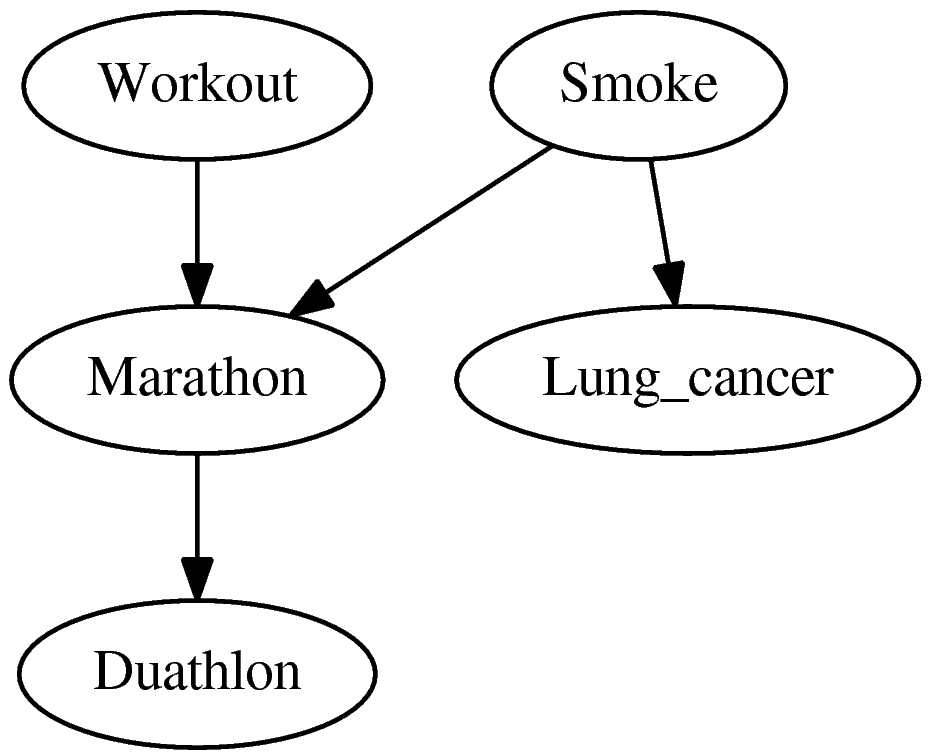}
  \end{subfigure}
  \begin{subfigure}{0.7\textwidth}
  \centering
    \includegraphics[scale=0.36]{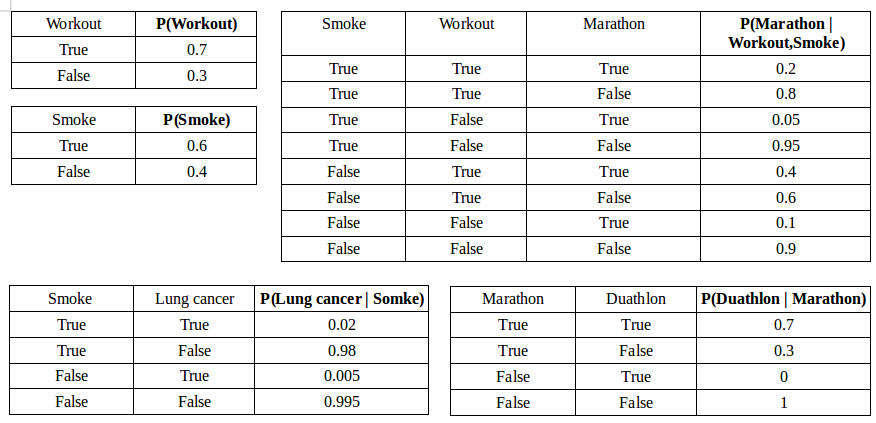}
  \end{subfigure}
  \caption{Bayesian Network over 5 variables.}
  \label{fig:bayes}
\end{figure}

These joint probabilities can be stored in a distribution table, but note that the dimension of the latter grows exponentially with the number of variables. This means that for most applications the table would be too large to be stored computationally. Alternatively, Bayesian networks, as in Fig.~\ref{fig:bayes}, allow for a compact representation of joint probability distributions \cite{Darwiche2008} as a directed acyclic graph structure. The advantage is that the space complexity of the representation can be made much smaller than in the general case, by exploiting conditional dependencies in the distribution, through the association to
 each graph node of a conditional probability table for each random variable, with directed edges representing conditional
dependencies.  For this reason, they are largely used in industrial applications. However, inference via a Bayesian Networks is still a  NP-problem. Fig.~\ref{fig:bayes} depicts a toy Bayesian network relating a few variables encoding different sorts of activities and the possibility of a lung cancer diagnosis.

\subsubsection{Inference.}
Algorithms that infer over a Bayesian network compute joint probabilities using the following equation:
\begin{equation} \label{eq:joint probability distribution}
	 P(x_1,...,x_n)=\prod_{i=1}^{n} P(x_i|Parents(Xi))
\end{equation}
The computational effort to determine this value is low because the number of values selected  is linearly bounded by the number of variables. The envisaged joint probability, however, may not be defined for all variables. If such is the case, it is necessary to sum out over all undefined variables as in
\begin{equation} \label{eq:Sumout}
	 P(X_1|x_2,x_3)=\sum_{x_4}\sum_{x_5} P(X_1,x_2,x_3,x_4,x_5)
\end{equation}
Consequently, the number of values to sum  out grows exponentially with the number of undefined variables. A well known algorithm for  variable elimination algorithm works exactly in this way. Approximate algorithms, treading off consumption of computational resources for precision, are typically used to tackle this problem. Solutions  are found faster but may not be precise. The literature documents a bunch of approximate algorithms. In this paper we will focus on \emph{rejection sampling} because the first part of the quantum inference algorithm discussed in the next section is a quantum analog to it. Rejection sampling is a popular method first systematised by  von Neumann \cite{Neu51}, who curiously enough also developed the Hilbert space formalization of quantum mechanics and its logic.

Rejection sampling generates samples resorting to the probability distributions defined by the conditional probability tables as depicted in Fig.~\ref{fig:bayes}. The consequence of this generation process is that a certain configuration of values for the variables is only sampled with the associated probability:

\begin{equation} \label{eq:Sampleprob}
	 P(Sample <X_1=true,X_2=false>)= P(X_1=true,X_2=false)
\end{equation}
Under those circumstances, a conditional probability can be determined by:
\begin{equation} \label{eq:Rejection}
	  P(X_1=true | X_2=false) \approx \frac{\# Samples (X_1=true, X_2=false)}{\# Samples (X_2=false)}
\end{equation}

\noindent
Clearly, the precision of the query grows with the number of useful samples ($\# Samples$). It is important to notice that not all samples are useful since samples with different values for the evidence variables are not used.

\subsubsection{Bayesian networks for decision making.}

    Bayesian networks are equipped with an \emph{utility function} in order to support decision-making processes. Its purpose  is to quantify the utility of possible outcomes. The \emph{expected utility} ($EU$) of an outcome is the product of its probability and the associated utility value. Formally, to find the expected utility of some action $a$, one computes
         \begin{equation} \label{eq:actuts}
        	 EU(a|e)= \sum_{r} P(Result=r|a,e)*U(r)
        \end{equation}
    If the $EU$ values of all feasible actions is known, it is possible to choose the `best', or more profitable, one:
        \begin{equation} \label{eq:actionsl}
        	  action = argmax_a EU(a|e)
        \end{equation}
    This is indeed  the maximum expected utility principle; a  rational entity is expected to  choose the action with the greatest expected utility with respect to her set of beliefs \cite{artificial}. 
    
    The previous principle describes many algorithms and solutions used in artificial intelligence. For example, in reinforcement learning a great number of \textit{agents} and  \textit{robots} are built on a process that attempts to find the optimal policy. This works with an instance of a Bayesian network (Markov decision process) and a more complex utility function (the so-called discounted reward function), where the agent also accounts for future rewards ($R(S_t)$), which are reachable from the starting state. Further, it values present rewards over future rewards with the use of a discount factor $ \gamma^t$.
    
    \begin{equation} \label{discounted}
        U_{\pi}(s)= E\left[\sum_{t=0}^{\infty} \gamma^t R(S_t)\right]
    \end{equation}

    \noindent Therefore, any algorithm that computes the best decision by Equations \eqref{eq:actuts} and \eqref{eq:actionsl} has the potential to be applied to all other instances derived from the principal one, as a consequence of having the same computational pattern.

\section{Quantum Bayesian inference}\label{sc:bayes}

\begin{figure}[H]
    \includegraphics[width=1\textwidth,height=0.28\textwidth]{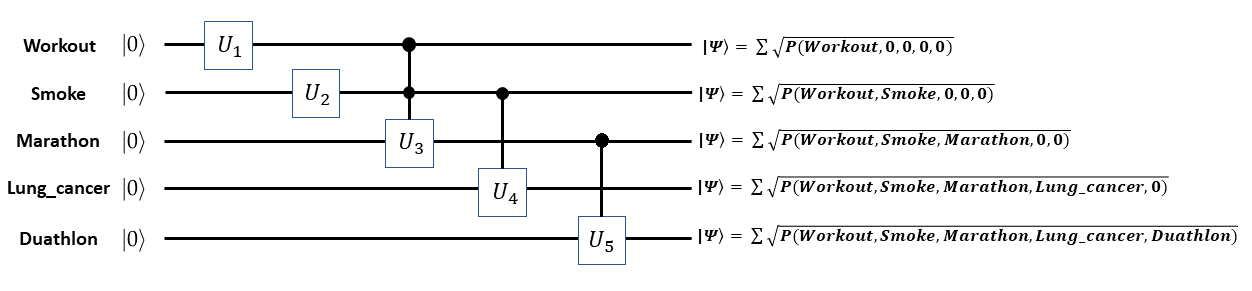}
  \caption{Encoding circuit.}
  \label{fig:circuito}
\end{figure}

In brief, a quantum algorithm can be regarded as a targeted manipulation of a quantum state (realised as an assembly of \emph{qubits}) with a subsequent measurement to retrieve relevant information.  States  are represented by column vectors of complex numbers whose sum of moduli squared is 1, often represented in the so-called Dirac notation \cite{NC10} as
$\ket{\Psi}$. Typically, they correspond to linear combinations of basis states affected by complex coeficients, as in, for example, equation \eqref{eq:State_variable}. The dynamics of a quantum system is represented by (the multiplication of the quantum state by) unitary matrices and is therefore reversible in time, as long as no measurement is involved. The reversal corresponds simply to a composition with the adjoint of the unitary matrix that represents forward evolution. Such an evolution can be expressed as a sequence of only a few elementary transformations represented as \emph{quantum gates}, which only act on one or two qubits at a time. Therefore, quantum algorithms are widely formulated as \emph{circuits} built of  these elementary gates, as depicted, for example, in Fig~\ref{fig:circuito}.

A quantum algorithm for inference on Bayesian networks was introduced by Low \emph{et al} \cite{Low2014}, which, as mentioned above, is based on an improved quantum version of the Rejection Sampling algorithm. This algorithm is able to generate samples quadratically faster than the classical version, provided that the network is not too densely connected. The algorithm is divided into 3 stages, detailed in the sequel.

In the first stage, the Bayesian network is encoded into a quantum state. For this, a binary variable can be represented by a single qubit and the probabilities are mapped to the coefficients of the quantum state:

\begin{equation} \label{eq:State_variable}
 \ket{\Psi} = \alpha \ket{Var_1=true} + \beta \ket{Var_1=false}  	\Leftrightarrow \ket{\Psi} = \begin{pmatrix}
     \alpha \\
     \beta
     \end{pmatrix}
\end{equation}

\noindent with the corresponding density matrix \cite{Barnett2009},

\begin{equation} \label{eq:measurement}
 \rho_{\Psi}=\ket{\Psi}\bra{\Psi}=\begin{pmatrix}
	   \alpha^2 & \alpha*\beta \\[4pt]
	   \alpha*\beta & \beta^2
	    \end{pmatrix}
\end{equation}

Whenever two variables share an edge in the network they are related, and therefore not independent from each other, such a relationship is expressed through state  entanglement. 
Entanglement represents a strong correlation between quantum states, therefore expressing shared information between different elements. 
The envisaged state is achieved though the application of  a specific sort of gates --- controlled rotations ---  to the state qubits.  The fact that a rotation is controlled by another qubit permits the creation of entanglement between them. The amplitude of the rotation defines the value of the coefficients. For instance, a circuit that encodes the Bayesian network in Fig.~\ref{fig:bayes} is represented in Fig~\ref{fig:circuito}.

\noindent
The whole quantum state is equivalent to a superposition of all on entries of the original joint probability distribution  table:

\begin{equation} \label{eq:whole_state}
\begin{split}
 \ket{\Psi'} = &\text{  } \gamma_1^2 \ket{Var_1=true,Var_2=true,...} + \gamma_2^2 \ket{Var_1=true,Var_2=false,...}  + \\ & \text{  } \gamma_3^2 \ket{Var_1=false,Var_2=true,..} + \gamma_4^2 \ket{Var_1=false,Var_2=false,...} + \\ & \text{  }...
\end{split}
\end{equation}
\noindent Afterward, a measurement\footnote{Notice, that the use of the density matrix notation with projectors is equivalent to the description of the measurement in the Dirac notation:
\begin{equation}
  \braket{Var_1=true,Var_2=true,...}{\Psi'} \equiv Tr(P_0 * \rho_{\Psi'})
\end{equation}
The matrix density notation was not required, as we are not dealing with mixed quantum states. However, it helped, later on, to expose the main ideas more clearly.} to this state produces a sample, as in the Rejection
Sampling algorithm, as the probability of each sample is the same as the one in
the distribution:

\begin{align*}
    P(Var_1=true,Var_2=true,...)=Tr(P_0 * \rho_{\Psi'})= \\
    \\ \begin{pmatrix}
	   1 & 0 & ...& 0 \\[4pt]
	   0 & 0 & ... & 0 \\[4pt]
	   \vdots & \vdots & \ddots & ... \\[4pt]
	  0 & 0 & ... & 0 
	    \end{pmatrix} * \begin{pmatrix}
	   \gamma_1^2 & ... & ...& ... \\[4pt]
	   ... & \gamma_2^2 & ... & ... \\[4pt]
	   ... & ... & \ddots & ... \\[4pt]
	   ... & ... & ... & \gamma_n^2 
	    \end{pmatrix}  = \gamma_1^2*1 + \gamma_2^2*0 + ... + \gamma_n^2*0 = \gamma_1^2 
\end{align*}

\noindent At this point, a quantum analog to Rejection Sampling is created. However, it is not an efficient way to do inference because every time we measure the state it collapses, and it is necessary to reconstruct the state entailing the need for a subsequent reconstruction.

In a second stage, the \textit{Amplitude Amplification} algorithm \cite{Brassard2000} is applied to amplify the states that have the right values for the evidence variable. It allows for a square root speed up in search problems, a fact that explains its relevance and ubiquity to many quantum programs. In our case, the quantum state that encodes the Bayesian network is divided into two orthogonal states, one where the evidence variables have the right value and another state where they lack it:

\begin{equation} 
\begin{split}
\label{eq:State_inic}
	\ket{\Psi_{init}}=&\sqrt{P(e)}\ket{Var_1,Var_2,...,evidences} \\ &+ \sqrt{1-P(e)}\ket{Var_1,Var_2,...,\neg{evidences}}
\end{split}
\end{equation}

\noindent
Next, the amplitude amplification algorithm is applied to search for the state that has the right values for the evidence variables \cite{Brassard2000}.

\begin{align*}
\label{eq:Grover}
 Q^k*\ket{\Psi_{init}} = &\cos{\left(\dfrac{2k+1}{2}*\theta\right)}\ket{Var_1,Var_2,...,evidences} \\ &+ \sin{\left(\dfrac{2k+1}{2}*\theta\right)}\ket{Var_1,Var_2,...,\neg{evidences}}
 \end{align*}

\noindent where $Q$ represent the operator that amplifies the selected elements, $k$ the number of iterations, and $\theta$ the initial probabilities,

\begin{equation}
\theta = 2sin^{-1}(\sqrt{P(e)}) \text{      }\wedge \text{      }\theta =  2cos^{-1}(\sqrt{1-P(e)})
\end{equation}

\noindent then for the right number of iteration ($k'$), the final quantum state approximates with great probability to the pretended state,

\begin{equation} 
\label{eq:Grover}
 Q^{k'}*\ket{\Psi_{init}} = \ket{\Psi_{final}} \approx\ket{Var_1,Var_2,...,evidences} 
\end{equation}

\noindent
The last stage amounts simply to observe this state and use the result as a sample. In Table~\ref{tab1} we can see the comparison between the classical and the quantum versions. The latter exhibits a quadratic speed up but only if the Bayesian network is not too densely connected, meaning that  $m$ the number of edges between the nodes ($n$) is not to large. Otherwise the price of encoding it to a quantum state will be too high, as the corresponding term grows exponentially. 

\begin{table}[h]
\centering
\caption{Classical \emph{vs} quantum complexity.}\label{tab1}
\begin{tabular}{|l|l|}
\hline
Process type & Complexity\\
\hline
Classical & $O(n*m*P(e)^-1)$  \\
\hline
Quantum & $O(n*2^m*P(e)^\frac{-1}{2})$\\
\hline 
\end{tabular}
\end{table}

\section{Quantum decision-making}\label{sc:dm}

Clearly, a quantum computer could be used to work out the conditional probabilities with a quadratic speed up for decision problems, according to equation \eqref{eq:actut}.

 \begin{equation} \label{eq:actut}
	 EU(a|e)= \sum_{r}{ \underbrace{P(Result=r|a,e)}_{Quantum}*\underbrace{U(r)}_{Classical}}
\end{equation}

In this section, however, we would like to propose a different approach which, in principle, will increase the advantage of having the quantum resources. The idea is  quite simple: Instead of sampling the conditional probabilities, the quantum state remains unobserved until the utility function is applied. The intention is to apply a transformation to the outcome variable and look to what happens to the action variable. As both the outcome and the action variables are entangled,  a transformation applied to the former will produce an effect on the latter.

The new algorithm modifies the process described  in the previous section to infer a conditional probability by preventing
 the action variable to be used as an evidence variable. Thus, after an application of the amplitude amplification algorithm and tracing out the non-evidence variables ($NE$), as nothing happens to them during the proposed process, we have,

    \begin{align} 
	  tr_{NE}&(Q_{Search_1}\ket{\Psi_{init}})=\ket{\Psi_{A,R}} = 
	  \gamma_{a,r} \ket{a,r,evidences} + \gamma_{a,\neg{r} } \ket{a,\neg{r} ,evidences} + \nonumber\\
	  & + \; \gamma_{\neg{a},r} \ket{\neg{a},r,evidences} +
	  \gamma_{\neg{a},\neg{r}} \ket{\neg{a},\neg{r},evidences}\label{eq:actiono}
	\end{align}
     
\noindent  or, equivalently,
        
	\begin{equation}
	     \ket{\Psi_{A,R}} = \begin{pmatrix}
	   \gamma_{a,r} \\
	   \gamma_{a,\neg{r} } \\
	    \gamma_{\neg{a},r} \\
	    \gamma_{\neg{a},\neg{r}} \\
	    \end{pmatrix}
	\end{equation}
 \noindent describing only the states of the featured variables. The utility function $U(r)$ is then applied to this state ($\ket{\Psi_{A,R}}$). Therefore, a quantum state $\ket{\Psi_{U}}$ isomorphic to the utility function will be created:
 
\begin{equation*}
\label{eq:action}
U(R)=\left\{
\begin{array}{ll}
U(r) \\
U(\neg{r}) \\
\end{array} 
\right.
\xrightarrow[\text{}]{\text{}}
\ket{\Psi_{U}} = 
\begin{pmatrix}
\dfrac{\sqrt{U(r)}}{n} \\[8pt]
\dfrac{\sqrt{U(\neg{r})}}{n} \\
\end{pmatrix} 
\end{equation*}

\noindent where $n$ is a normalization term such that $\bra{\Psi_{A,R}}\ket{\Psi_{A,R}}$ sums up to 1. Also, by creating both states in memory, the whole product state $\ket{\Psi_{A,R}} \otimes \ket{\Psi_{U}}$ becomes,

 \begin{equation}
     \ket{\Psi_{Syt}} = \ket{\Psi_{A,R}} \otimes \ket{\Psi_{U}}=  \begin{pmatrix}
	    \gamma_{a,r} \\
	    \gamma_{a,\neg{r} } \\
	    \gamma_{\neg{a},r} \\
	    \gamma_{\neg{a},\neg{r}} \\
	    \end{pmatrix} 
	    \otimes 
	    \begin{pmatrix}
	    \dfrac{\sqrt{U(r)}}{n} \\[8pt]
	    \dfrac{\sqrt{U(\neg{r})}}{n} \\
	    \end{pmatrix} =
	    \begin{pmatrix}
	    \gamma_{a,r} * \dfrac{\sqrt{U(r)}}{n} \\[8pt]
	    
	    \gamma_{a,r} * \dfrac{\sqrt{U(\neg{r})}}{n} \\[8pt]
	    
	    \gamma_{a,\neg{r} } *\dfrac{\sqrt{U(r)}}{n} \\[8pt]
	    
	    \gamma_{a,\neg{r} } *\dfrac{\sqrt{U(\neg{r})}}{n} \\[8pt]
	    
	    \gamma_{\neg{a},r} * \dfrac{\sqrt{U(r)}}{n} \\[8pt]
	    
	    \gamma_{\neg{a},r} * \dfrac{\sqrt{U(\neg{r})}}{n} \\[8pt]
	    
	    \gamma_{\neg{a},\neg{r}} * \dfrac{\sqrt{U(r)}}{n}\\[8pt]
	    
	    \gamma_{\neg{a},\neg{r}} * \dfrac{\sqrt{U(\neg{r})}}{n}\\
	    \end{pmatrix} 
 \end{equation}

\noindent Then, this state $\ket{\Psi_{Syt}}$ already contains the relevant terms where the Utility function is applied to the correct bases. All one has to do is to amplify them resorting again to amplitude amplification algorithm. For the example at hands, such is the case when $r \land r$ and $\neg{r} \land \neg{r}$ hold, yielding,

 \begin{equation}
     Q_{Search_2}\ket{\Psi_{Syt}} \approx 
      \begin{pmatrix}
	    \dfrac{ \gamma_{a,r} * \sqrt{U(r)}}{n'} \\[8pt]
	    
	    0 \\[8pt]
	    
	    0 \\[8pt]
	    
	    \dfrac{\gamma_{a,\neg{r} } * \sqrt{U(\neg{r})}}{n'} \\[8pt]
	    
	    \dfrac{\gamma_{\neg{a},r} * \sqrt{U(r)}}{n'} \\[8pt]
	    
	    0 \\[8pt]
	    
	    0 \\[8pt]
	    
	     \dfrac{\gamma_{\neg{a},\neg{r}} * \sqrt{U(\neg{r})}}{n'}\\
	    \end{pmatrix} 
 \end{equation}
 
 \noindent At this moment the amplitudes of the action variable hold the solution to the decision problem. To understand how the Results variable and the Utility function are traced out as follows,

 \begin{align}
     \rho_A & =t r_{R,U}(\rho_{Syt}) = \\
     & \begin{pmatrix}
	    (\dfrac{ \gamma_{a,r} * \sqrt{U(r)} + \gamma_{a,\neg{r} } * \sqrt{U(\neg{r})}}{n'})^2  & ... \\[8pt]
	    ... & (\dfrac{\gamma_{\neg{a},r} * \sqrt{U(r)} + \gamma_{\neg{a},\neg{r}} * \sqrt{U(\neg{r})}}{n'})^2  \nonumber\\
	    \end{pmatrix}
 \end{align}
 
 \noindent A measurement yields,
 
 \begin{equation}
     P(a')= tr(P_0 * \rho_{Syt})=
     \begin{pmatrix}
     1 &  0 \\
     0 &  0
     \end{pmatrix} *  \rho_{Syt} =  (\dfrac{ \gamma_{a,r} * \sqrt{U(r)} + \gamma_{a,\neg{r} } * \sqrt{U(\neg{r})}}{n'})^2 
 \end{equation}
 
  \noindent and,
 
  \begin{equation}
     P(\neg{a'})= tr(P_1 * \rho_{Syt})=
     \begin{pmatrix}
     0 &  0 \\
     0 &  1
     \end{pmatrix} *  \rho_{Syt} =   (\dfrac{\gamma_{\neg{a},r} * \sqrt{U(r)} + \gamma_{\neg{a},\neg{r}} * \sqrt{U(\neg{r})}}{n'})^2
 \end{equation}

\noindent Combining the deduced probability of the action variable with,
	
	\begin{equation} \label{eq:action5}
	  \gamma_{a,r}^2 = P(a,r,evidences) 
	\end{equation}
	and
	\begin{equation} \label{eq:action6}
	 P(r|a,evidences)= \dfrac{P(a,r,evidences)}{P(a,evidences)}
	\end{equation}
	we  conclude that
	
	\begin{equation} \label{eq:action7}
	  \gamma_{a,r}^2 \propto P(r|a,evidences)
	\end{equation}
	
	\noindent
	and the transformation yields a state where
	
	\begin{equation}\begin{split} \label{eq:action8}
	 P(a_i')=  tr(P_i * \rho_{Syt}) = \dfrac{ \gamma_{a_i,r_1}^2 * U(r_1) + \gamma_{a_i,r_2 }^2 * U(r_2) + ... + \gamma_{a_i,r_n}^2 * U(r_n) }{n'^2}
	\end{split}
	\end{equation}
	
	\noindent So, after applying there transformations the probability of some action ($a_i$) is proportional to its expected utility (Equation \ref{eq:actuts}),

	\begin{equation} \label{eq:action9}
	  P(a_i') \propto EU(a_i|e)
	\end{equation}
	
	\begin{equation} \label{eq:action10}
	  P(a_i') = \varrho_{a_i} * EU(a_i|e)
	\end{equation}
	
	\noindent
	  Finally, if initially the Bayesian networks respects,
	
	 \begin{equation} \label{eq:independence}
	  P(a| evidences) = P(a)
	\end{equation}

	\noindent then
	 \begin{equation} \label{eq:action11}
	  P(r|a,evidences) = \dfrac{P(r,a|evidences)}{P(a)}
	\end{equation}

	\noindent and
	\begin{equation} \label{eq:actiosn}
	 P(a_{i'})=P(a_i), i\neq i'
	\end{equation}
	
	\noindent The constant of proportionality takes the following value for all actions:

	\begin{equation}
	 \varrho_{i'} = \varrho_i = \dfrac{1}{n'^2*P(a)} \text{ , } i\neq i'
	\end{equation}
	
\begin{figure}[H]
            \centering
         \includegraphics[scale=0.35]{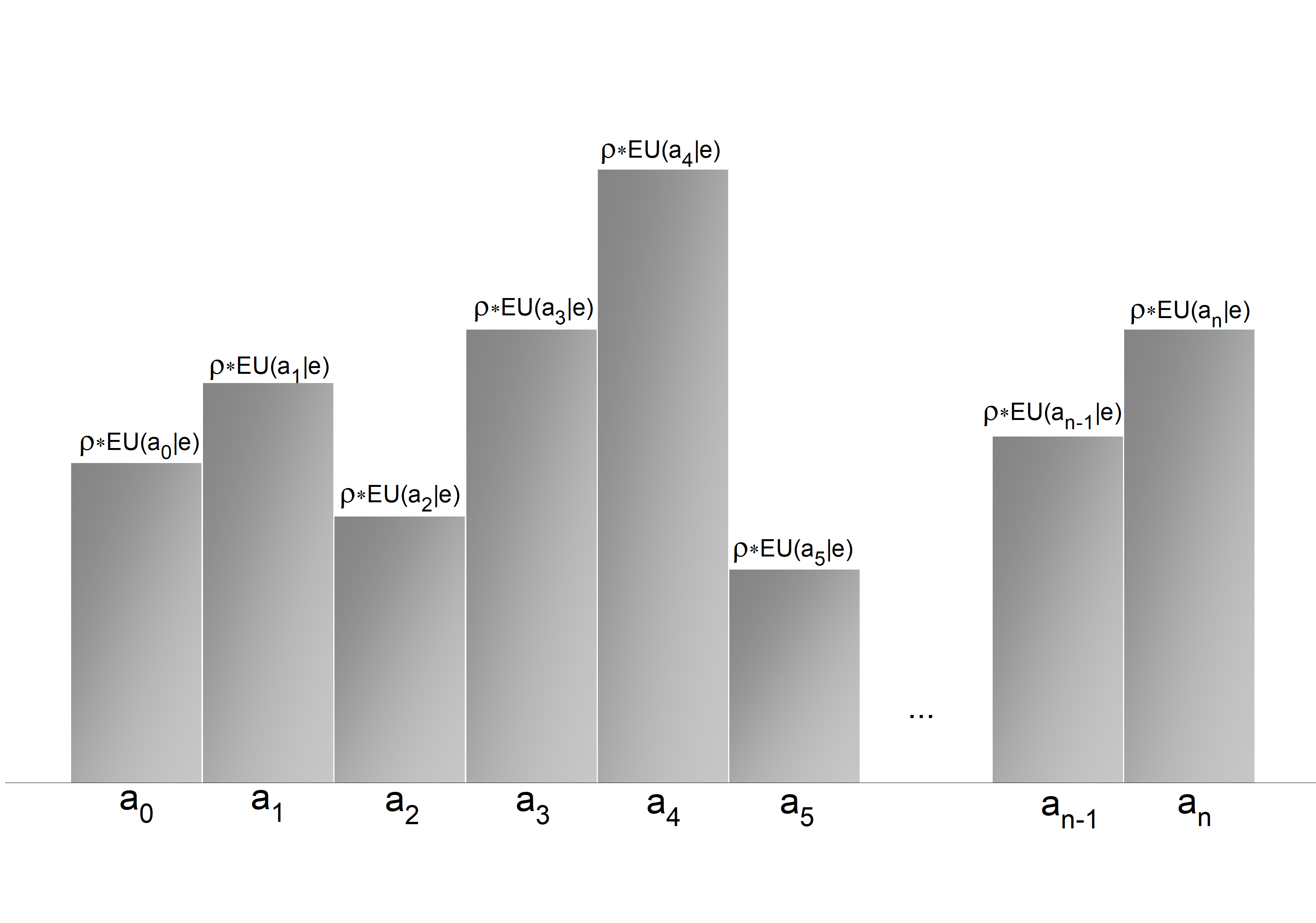}
            \caption{Probability distribution of an action variable.}
            \label{fig:grafico}
        \end{figure}

	\noindent Thus, the values of the proportionality constants $\varrho_i$ between all the Expected Utilities 
        remain the same. This means that an action with a greater probability has a greater expected utility. Consequently,  to  choose  the  action  with  the  greatest  probability/utility  (Figure~\ref{fig:grafico}),  it is  enough  to  resort  to  a  limited  collection  of  samples,  rather  than  obtaining  first  all  the conditional probabilities. Moreover, this provides a more precise way of choosing an action because the sampling method always yields an approximation and the error affecting the computed values grows for every conditional probability determined. 

\begin{figure}[H]
	    \centering
	    \includegraphics[scale=0.35]{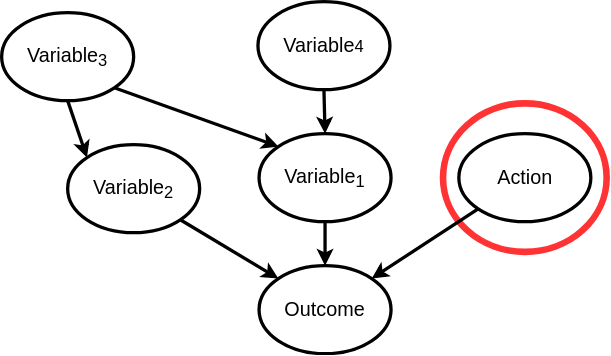}
	    \caption{Bayesian network with an independent action node.  }
	    \label{fig:topo}
	\end{figure}

        	Remember, that this decision process requires that Equation~\eqref{eq:actiosn} has to be initially true. This expresses the rational choice which considers all actions as equal
	at the beginning. In other words, the intelligent agent is not biased beforehand. Additionally, the action variable should be independent of the evidence
	variables as in Equation \eqref{eq:independence}, which means that the topology of the network has to be as in Figure~\ref{fig:topo}. This requirement ensures that the intelligent agent 
	is not biased by the current state of his environment and performs his decisions in order to achieve the best outcome in the future state. 
	
\section{Complexity} \label{cplx}
 The purpose of this Section is to characterize computational complexity of  the proposed algorithm and compare it to the solution that computes the conditional probabilities with use of the quantum inference algorithm (Equation \eqref{eq:actut}). So, to simplify let us denote by Process A the new quantum algorithm and by Process B the second one.
 
 Both algorithms generate samples to determine which is the best action. The number of operations ($I_t$) in each algorithm is defined by the number of iterations
	per sample ($I_s$) and the number of samples ($S$) necessary, as in Equation \eqref{eq:iterations}. 
	
	\begin{equation} \label{eq:iterations}
	  I_t=S*I_s
	  \end{equation}
	  
	  \subsubsection{Number of iterations}
	  
	  The number of iterations per sample  of the two processes are defined by the number of \textit{Search} iterations that are necessary to apply in each case. Also, the number of iterations necessary
	to find the goal state in a quantum search is defined by the probability of this state:
	
	\begin{equation} \label{eq:comp1}
	 I_s=\sqrt{\frac{1}{P(state)}}
	\end{equation}
	
	For Process A, this is the probability of the state which has already the utility function applied to it,
	
	\begin{equation} \label{eq:comp2}
	 P(state)=\sum_r U(r)* P(r,e)
	\end{equation}
	
	\noindent knowing that,

	\begin{equation} \label{eq:comp3}
	1=\sum_r U(r)
	\end{equation}

		\noindent Assuming that any  distribution is possible for U(r) and P(r,e),  we conclude that the probability can take any value between $0$ and P(e). We also know that the mean value for $U(r)$ is:

	\begin{equation}
	    \frac{1}{N_r}
	\end{equation}

	\noindent where $N_r$ represents the dimension of the outcome variable. Thus, P(e,r) can be described as:

	\begin{equation}
	    \frac{P(e)}{N_r}
	\end{equation}

	\noindent So the mean value for the product of the two values $U(r)*P(e,r)$ will be,

	\begin{equation}
	    \frac{P(e)}{N_r}*\frac{1}{N_r}=\frac{P(e)}{N_r^2}
	\end{equation}

	\noindent if they are independent, which is the case because the utility function is independent of the information present in the Bayesian Network. 
	The mean value for the sum can be computed by the sum over the mean terms

	\begin{equation}
	    Mean(P(state))=\sum_r \frac{P(e)}{N_r^2} = \frac{P(e)*N_r}{N_r^2} = \frac{P(e)}{N_r}
	\end{equation}
	
	\noindent This mean value for the probability will be used to define the number of steps:

	\begin{equation} \label{eq:comp5}
	  I_s=\sqrt{\frac{N_r}{P(e)}}
	\end{equation}
	
	\noindent defining in this way the number of iterations necessary to obtain a sample with Process A. 
	
	The same has to be done for Process B, where the probabilty of the goal state is
	
	\begin{equation} \label{eq:comp6}
	   P(state)= P(e,a)
	\end{equation}

	\noindent In this case, we have to apply the requirements determined by Process A described in \eqref{eq:independence} and \eqref{eq:actiosn} in order to make a correct comparison at a later stage,

	  \begin{equation} \label{eq:comp7}
		P(e,a)=P(e)*P(a)=\frac{P(e)}{N_a}
		\end{equation}

	\noindent where $N_a$ is the dimension of the action variable. Finally, we estimate the number of iterations as

	\begin{equation} \label{eq:comp8}
	 I_s=\sqrt{\frac{N_a}{P(e)}}
	\end{equation}

	\subsubsection{Number of samples}
	
	The next step is to obtain the number of samples necessary for each process. Recall that the simultaneous error terms for a \textit{Multinomial Distribution} are:
	\begin{equation}
	    (pi-\pi_i)^2= \frac{A*\pi_i(1-\pi_i)}{N},  (i=1,2,...,k)
	\end{equation}

\begin{figure}[H]
	    \centering
	    \includegraphics[scale=0.3]{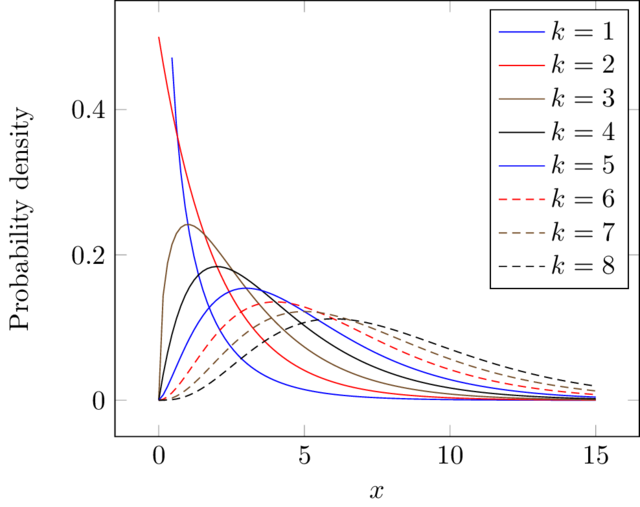}
	    \caption{Chi-square distributions with different degrees of freedom ($k$).}
	    \label{fig:chi}
	\end{figure}

	The value A represents the upper $\alpha*100-th$ percentile of a \textit{Chi-Square Distribution} (figure~\ref{fig:chi}) with k-1 degrees of freedom, $\pi_i$ represent the probability of category i,
	$N$ is the number of samples, and the difference on the left side of the equation represents the error term \cite{Goodman1965}. 
	
	\noindent Writing the same equation as a  function of $N$, yields

	\begin{equation}\label{samples}
	    N= \frac{A*\pi_i(1-\pi_i)}{\delta^2},  (i=1,2,...,k)
	\end{equation}
	
	Equation \eqref{samples}, defines the number of samples necessary for Process A and Process B, since, both processes are sampling from a quantum state with
	multiple bases.

	\subsubsection{Total number of operations}
	
	The total number of operations is characterized by the product of the terms deduced in the previous sections and the number of operations necessary to encode the 
	network as a quantum state. Additionally, Process B requires at least $2(N_a+N_r)$ operations to apply the Utility function and sum the respective terms for the expected utilities. Finally, the total number of operations for each process to solve the decision problem is shown in Table~\ref{tab4}. 
	
	 \begin{table}[H]
	      \centering
	      \caption{Mean number of operations for each process.}\label{tab4}
	      \begin{tabular}{|c|c|c|}
	      \hline
	      Process A & Process B\\
	      \hline
         $n*2^{m}*\sqrt{\frac{N_r}{P(e)}}*\frac{A*\pi_i(1-\pi_i)}{\delta_a^2}$  & $n*2^m*\sqrt{\frac{N_a}{P(e)}}*\frac{A*\pi_i(1-\pi_i)}{\delta_c^2} *N_a  + 2(N_a+N_r)$ \\
	      \hline
	      \end{tabular}
	      \end{table}

 When the decision problem is totally defined all that it requires is  to plug the number in the equation and look which process performs better. However, a comparison was performed (as described in Appendix~\ref{comp}) and the relation between the computational effort is asymptotically over the term,
  
  	\begin{equation}
	   \dfrac{Process B}{Process A}  \geq \sqrt{\frac{N_r}{N_a}}
	\end{equation}
  
  This result shows that Process A is faster when the outcome variable has a greater dimension than the action variable. This is a quite normal scenario in real applications because the number of states in which an agent can transit  is  tremendously  smaller  than  the  possible  states  that  his  environment can evolve. Also, for a fixed number of action this process allows the agent to explore quadratically more outcomes with the same computational effort, making him a \emph{wiser} decision maker.
  
  Finally, this process can also be compared with  a quantum version of \emph{decision networks} \cite{artificial}, as discussed in \cite{Oliveira2019}. The results show that again the best process depends on the characteristics of the problem. Although, it is important to mention that process A solves a decision process that wants to sample an action from a distribution based on the expected utilities  in a extremely efficient way with only \emph{one sample}, this kind of decision processes could be used and studied for applications in reinforcement learning.

	\section{Proof-of-concept implementation}\label{poc} \label{implementation}
	
	 The algorithm presented in Section~\ref{sc:dm} was implemented on the IBM Q quantum simulator as a proof-of-concept. At our disposal was the IBM 20-qubit machine, which is based on superconducting circuits \cite{Steffen2011}. This machine specifies an error term associated with each gate used in a
quantum circuit and a life-time for each qubit. So, as the number of gates grows the error of the outcome grows as well. The output of a circuit with a considerable number of gates would be majorly noise. Thus, the decision processes presented before, which is based on a search problem, would be impossible to compute with a manageable error term.

IBM’s best quantum computer is not the only that fails to solve such problems. The best quantum
devices, in the world, are not even near to solve problems related to search problems with a higher dimension. However, that does not mean that the current devices are completely useless. There are
problems where a \textit{Noisy Intermediate-Scale Quantum} (NISQ) devices may have an impact, in the near future \cite{Preskill2018}. The applications of these NISQ devices are
related to simulations in chemistry and many-body quantum physics. \footnote{It is interesting to mention that the major companies investing in quantum computing are constructing devices based on different technologies. Microsoft devices are based on topological quantum computing \cite{Nayak2008}, while Intel is exploring spin qubits \cite{Vandersypen2017}.}

\begin{figure}

  \begin{subfigure}{0.4\textwidth}
  \centering
   \includegraphics[scale=0.5]{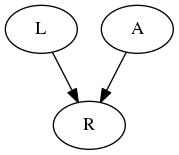}
  \end{subfigure}
  \begin{subfigure}{0.6\textwidth}
  \centering
    \includegraphics[scale=0.5]{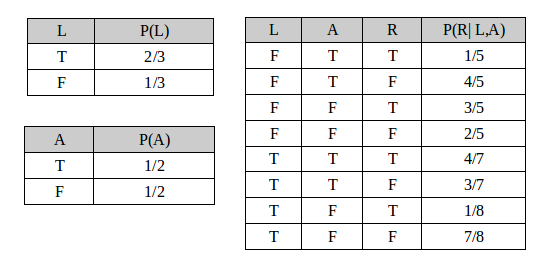}
  \end{subfigure}
 \caption{Bayesian network over 3 variables. Node L represents the evidence variable, A the action variable and R the outcome variable.}
\label{fig:Rede_Bayes_ibm}
\end{figure}

Over the last years, quantum devices have had a lot of progress. For example, the number of qubits are smoothly increasing, the gate errors are reducing \cite{Schafer2018} and entanglement between them is becoming stronger \cite{Kues2017,Pirandola2006}. This progress has been giving hope to construct a powerful universal quantum computer, which one day may have a great impact on our everyday life. But to validate results as pretended, in this section a classical simulator has to be used. However, the same simulator struggles to compute the outcomes, if the number of qubits used increases. As mentioned before the complexity to simulate a quantum computer on a classical computer is too high. Given that, a very simple Bayesian network (Figure~\ref{fig:Rede_Bayes_ibm}) was selected for the decision process.

\begin{figure}[H]
	\centering
	    \includegraphics[scale=0.5]{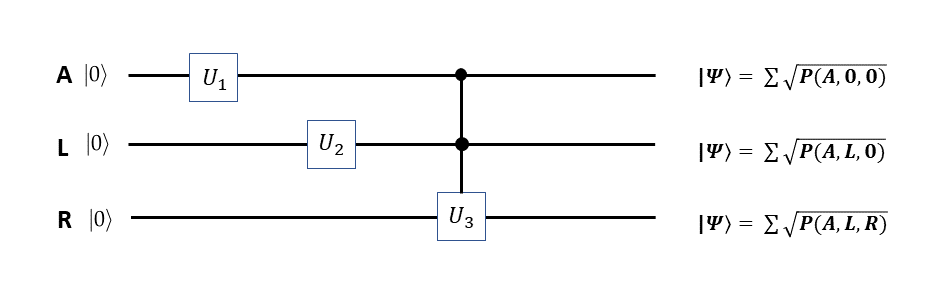}
	  \caption{Quantum circuit composed by rotations and controlled-rotations.}
	  \label{fig:circuitob}
	\end{figure}

	\noindent The network was encoded to a quantum state with use of the technique presented in \cite{Low2014}, producing the circuit shown in Figure~\ref{fig:circuitob}.

\begin{figure}[H]
   \begin{subfigure}[c]{0.45\textwidth}
     \centering
     \includegraphics[scale=0.7]{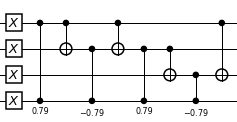}
   \end{subfigure}
  {\LARGE$\ \ \ \ \ \ \ \  ...$}%
   \begin{subfigure}[c]{0.45\textwidth}
     \centering
     \includegraphics[scale=0.7]{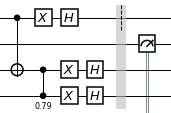}
   \end{subfigure}
   \caption{Circuit representation of the amplitude amplification algorithm for the decision-making process.}
  \label{fig:gdecomp}
\end{figure}

	A value for the evidence variable $L$ was selected ($L=False)$ and the following utility function,
	\begin{equation} \label{eq:action3}
		      U(R)=\left\{
		      \begin{array}{ll}
		      \ 7\ \ \ \ , R=r \\
		      \ 3\ \ \ \ , R= \neg{r}\\
		      \end{array} 
		      \right.
		       \end{equation}
		       
    \noindent was applied with use of the proposed algorithm. However, to compute the algorithm on IBM’s quantum simulator, each part has be to converted in a concrete quantum circuit (Figure~\ref{fig:gdecomp}). Every state has to be encoded, which in theory is simple with the use of rotations and controlled-rotations. The major difficulty exists when the rotation is controlled by more than one qubit. In such cases, this operation has to be decomposed to simpler and available operations in the working framework. The existence of an equivalent circuit is guaranteed by the fact that the simulator is a universal quantum computer, meaning that it may perform any possible computation. In practice, there are tools to decompose complex operations into simpler ones \cite{Vartiainen2004,Mottonen2004}.

	  Finally, these circuits have to be built on "Qiskit" which runs on a "jupyter notebook" (Github link \href{https://github.com/MichaelOliveira1994/Quantum-Bayesian-Decisions}{"Quantum
Bayesian Decisions"}). The circuits created are sent as a job to IBM’s servers and the results are  sent  back  to  the  client. For this example, a significant number of samples was generated. The number of samples for each state of the action variable must be similar to the theoretical probability. Table~\ref{tab2} shows that this is indeed the case: the experimental result is quite similar to the one  foreseen by theory.

	      The small discrepancies pointed out in Table~\ref{tab2} can be explained by deficiencies of the implementation. 
	      First, note that the amplitude amplification algorithm is probabilistic, i.e. the result is never entirely precise.
	      On the other hand, the number of iterations in the amplitude amplification algorithm was an integer number; thus, if $n.m$ non-integer iterations are required the
	      usage of the value bellow $n$ or the one above $n+1$, generates a small variation. Actually, it is possible to perform a a quantum search with a non-integer 
	      number of iterations \cite{Zekrifa2000}, but this would not add relevance to this results.

	      \begin{table}[H]
	      \centering
	      \caption{Comparison between theoretical and experimental results after sampling from the constructed state.}\label{tab2}
	      \begin{tabular}{|c|c|c|}
	      \hline
	      States & Theoretically expected  probability & Percentage of Samples\\
	      \hline
	      $Action_0$ & 0,58 & 0.544 \\
	      \hline
	      $Action_1$ & 0,42 & 0.456 \\
	      \hline 
	      \end{tabular}
	      \end{table}

	   \section{Conclusions and Future work} \label{conc}
   A quantum algorithm which solves the generic decision-making  problem was presented. It is a very curious solution for couple of reasons. First, it has a proven computational advantage over the classical and the semi-classical solutions when the parameters are in the correct relation. Secondly, it samples from a very particular probability distribution, which classically would require an tremendous amount of  computational work to recreate. Moreover, it benefits from the structure of the data, which  no  classical  algorithm  could  benefit  from, making it a very interesting example to illustrating the differences between classical and quantum computations. 
   
   To support the theoretical work described the algorithm was implemented in IBM's quantum simulator as a proof-of-concept. The results were as in correspondence with what theory anticipated for the example chosen, further confirming the ideas presented.
  
    As an extension to this work we propose a search for decision problems that take advantage by sampling from the probability distribution created by this solution. Also, the decision-making process discussed here was related to a static model, in which, neither the utility function nor the Bayesian network change in time. It would be of interest to verify if the decision-making process could benefit from an additional learning process \cite{Jonsson2007,Robinson2010,Tong2001}. Enabling an agent to adapt its behaviour to a changing environment, would probably result in better outcomes, raising the number of possible applications.

%
%
\bibliographystyle{splncs04}
\bibliography{library}

\begin{thebibliography}{10}
\providecommand{\url}[1]{\texttt{#1}}
\providecommand{\urlprefix}{URL }
\providecommand{\doi}[1]{https://doi.org/#1}

\bibitem{Al-Jarrah20155}
Al-Jarrah, O.Y., Yoo, P.D., Muhaidat, S., Karagiannidis, G.K., Taha, K.:
  {Efficient Machine Learning for Big Data: Review}. CoRR  \textbf{abs/1503.0}
  (2015), \url{http://arxiv.org/abs/1503.05296}

\bibitem{Barnett2009}
Barnett, S.: {Quantum Information}. Oxford University Press, Inc., USA (2009)

\bibitem{Biamonte2016}
Biamonte, J., Wittek, P., Pancotti, N., Rebentrost, P., Wiebe, N., Lloyd, S.:
  {Quantum Machine Learning}. Nature  \textbf{549} (2016).
  \doi{10.1038/nature23474}

\bibitem{Brassard2000}
Brassard, G., Hoyer, P., Mosca, M., Tapp, A.: {Quantum Amplitude Amplification
  and Estimation}. arXiv e-prints pp. quant--ph/0005055 (2000)

\bibitem{Darwiche2008}
Darwiche, A.: {Chapter 11 Bayesian Networks}. Foundations of Artificial
  Intelligence  \textbf{3}(07),  467--509 (2008).
  \doi{10.1016/S1574-6526(07)03011-8}

\bibitem{Giovannetti2008}
Giovannetti, V., Lloyd, S., Maccone, L.: {Quantum Random Access Memory}.
  Physical review letters  \textbf{100},  160501 (2008).
  \doi{10.1103/PhysRevLett.100.160501}

\bibitem{Goodman1965}
Goodman, L.A.: {On Simultaneous Confidence Intervals for Multinomial
  Proportions}. Technometrics  \textbf{7}(2),  247--254 (1965).
  \doi{10.1080/00401706.1965.10490252},
  \url{https://amstat.tandfonline.com/doi/abs/10.1080/00401706.1965.10490252}

\bibitem{Harrow2009}
Harrow, A.W., Hassidim, A., Lloyd, S.: {Quantum Algorithm for Linear Systems of
  Equations}. Phys. Rev. Lett.  \textbf{103}(15),  150502 (2009).
  \doi{10.1103/PhysRevLett.103.150502},
  \url{https://link.aps.org/doi/10.1103/PhysRevLett.103.150502}

\bibitem{Inglot2010}
Inglot, T.: {Inequalities for quantiles of the chi-square distribution}.
  Probability and Mathematical Statistics  \textbf{30} (2010)

\bibitem{Jonsson2007}
Jonsson, A., Barto, A.: {Active Learning of Dynamic Bayesian Networks in Markov
  Decision Processes}. In: Proceedings of the 7th International Conference on
  Abstraction, Reformulation, and Approximation. pp. 273--284. SARA'07,
  Springer-Verlag, Berlin, Heidelberg (2007),
  \url{http://dl.acm.org/citation.cfm?id=1770681.1770705}

\bibitem{Kues2017}
Kues, M., Reimer, C., Roztocki, P., Cort{\'{e}}s, L.R., Sciara, S., Wetzel, B.,
  Zhang, Y., Cino, A., Chu, S.T., Little, B.E., Moss, D.J., Caspani, L.,
  Aza{\~{n}}a, J., Morandotti, R.: {On-chip generation of high-dimensional
  entangled quantum states and their coherent control}. Nature  \textbf{546},
  ~622 (jun 2017), \url{https://doi.org/10.1038/nature22986
  http://10.0.4.14/nature22986}

\bibitem{Li2018}
Li, C., Welling, M., Zhu, J., Zhang, B.: {Graphical Generative Adversarial
  Networks}. CoRR  \textbf{abs/1804.0} (2018),
  \url{http://arxiv.org/abs/1804.03429}

\bibitem{Low2014}
Low, G.H., Yoder, T.J., Chuang, I.L.: {Quantum inference on Bayesian networks}.
  Phys. Rev. A  \textbf{89}(6),  62315 (jun 2014).
  \doi{10.1103/PhysRevA.89.062315},
  \url{https://link.aps.org/doi/10.1103/PhysRevA.89.062315}

\bibitem{Man09}
Mansinghka, V.K.: {Natively probabilistic computation}. Ph.D. thesis,
  Massachusetts Institute of Technology (2009)

\bibitem{Mottonen2004}
M{\"{o}}tt{\"{o}}nen, M., Vartiainen, J.J., Bergholm, V., Salomaa, M.M.:
  {Quantum Circuits for General Multiqubit Gates}. Phys. Rev. Lett.
  \textbf{93}(13),  130502 (2004). \doi{10.1103/PhysRevLett.93.130502},
  \url{https://link.aps.org/doi/10.1103/PhysRevLett.93.130502}

\bibitem{Nayak2008}
Nayak, C., Simon, S.H., Stern, A., Freedman, M., {Das Sarma}, S.: {Non-Abelian
  anyons and topological quantum computation}. Rev. Mod. Phys.  \textbf{80}(3),
   1083--1159 (2008). \doi{10.1103/RevModPhys.80.1083},
  \url{https://link.aps.org/doi/10.1103/RevModPhys.80.1083}

\bibitem{Neal1996}
Neal, R.M.: {Bayesian Learning for Neural Networks}. Springer-Verlag, Berlin,
  Heidelberg (1996)

\bibitem{Neu51}
von Neumann, J.: {Various techniques used in connection with random digits}.
  Monte Carlo Method, Appl. Math. Series pp. 36--38 (1951)

\bibitem{NC10}
Nielsen, M.A., Chuang, I.L.: {Quantum Computation and Quantum Information (10th
  Anniversary Edition)}. Cambridge University Press (2010)

\bibitem{Oliveira2019}
Oliveira, M.: {On Quantum Bayesian Networks}. Physical engineering, University
  of Minho (2019)

\bibitem{OzolsRR12}
Ozols, M., Roetteler, M., Roland, J.: {Quantum rejection sampling}. In:
  Innovations in Theoretical Computer Science 2012, Cambridge, MA, USA, January
  8-10, 2012. pp. 290--308. ACM (2012)

\bibitem{Pirandola2006}
Pirandola, S., Vitali, D., Tombesi, P., Lloyd, S.: {Macroscopic Entanglement by
  Entanglement Swapping}. Phys. Rev. Lett.  \textbf{97}(15),  150403 (2006).
  \doi{10.1103/PhysRevLett.97.150403},
  \url{https://link.aps.org/doi/10.1103/PhysRevLett.97.150403}

\bibitem{Preskill2018}
Preskill, J.: {Quantum {\{}C{\}}omputing in the {\{}NISQ{\}} era and beyond}.
  Quantum  \textbf{2}, ~79 (2018). \doi{10.22331/q-2018-08-06-79},
  \url{https://doi.org/10.22331/q-2018-08-06-79}

\bibitem{Robinson2010}
Robinson, J.W., Hartemink, A.J.: {Learning Non-Stationary Dynamic Bayesian
  Networks}. J. Mach. Learn. Res.  \textbf{11},  3647--3680 (2010),
  \url{http://dl.acm.org/citation.cfm?id=1756006.1953047}

\bibitem{artificial}
Russel, S., Norvig, P.: {Artficial Intelligence : a modern approach}. Upper
  Saddle River, NJ : Prentice Hall, ed, 3rd edn. (2010)

\bibitem{Sakkaris2016}
Sakkaris, P.: {QuDot Nets: Quantum Computers and Bayesian Networks}. arXiv
  e-prints  (2016)

\bibitem{Schafer2018}
Sch{\"{a}}fer, V.M., Ballance, C.J., Thirumalai, K., Stephenson, L.J.,
  Ballance, T.G., Steane, A.M., Lucas, D.M.: {Fast quantum logic gates with
  trapped-ion qubits}. Nature  \textbf{555}, ~75 (feb 2018),
  \url{https://doi.org/10.1038/nature25737 http://10.0.4.14/nature25737}

\bibitem{Steffen2011}
Steffen, M., DiVincenzo, D.P., Chow, J.M., Theis, T.N., Ketchen, M.B.: {Quantum
  computing: An IBM perspective}. IBM Journal of Research and Development
  \textbf{55}(5),  13:1--13:11 (2011). \doi{10.1147/JRD.2011.2165678}

\bibitem{Tong2001}
Tong, S., Koller, D.: {Active Learning for Parameter Estimation in Bayesian
  Networks}. Proc 13th Conf Neural Information Processing  (2001)

\bibitem{Vandersypen2017}
Vandersypen, L.M.K., Bluhm, H., Clarke, J.S., Dzurak, A., Ishihara, R.,
  Morello, A., Reilly, D.J., Schreiber, L.R., Veldhorst, M.: {Interfacing spin
  qubits in quantum dots and donors—hot, dense, and coherent}. npj Quantum
  Information  \textbf{3},  1--10 (2017)

\bibitem{Vartiainen2004}
Vartiainen, J.J., M{\"{o}}tt{\"{o}}nen, M., Salomaa, M.M.: {Efficient
  Decomposition of Quantum Gates}. Phys. Rev. Lett.  \textbf{92}(17),  177902
  (2004). \doi{10.1103/PhysRevLett.92.177902},
  \url{https://link.aps.org/doi/10.1103/PhysRevLett.92.177902}

\bibitem{Zekrifa2000}
Zekrifa, D.M.S., Hoyer, P., Mosca, M., Tapp, A.: {Quantum Amplitude
  Amplification and Estimation}. AMS Contemporary Mathematics Series
  \textbf{305} (2000). \doi{10.1090/conm/305/05215}

\end{thebibliography}

	\appendix
	
	\section{Complexity comparison}\label{comp}

The decision-making processes we aim at comparing require inequality \eqref{eq:aciones} to  be  satisfied.   It  assures  that  the  decision  maker  chooses  with certainty the best action.

    	\begin{equation} \label{eq:aciones}
	\forall_{n\setminus \{max\}}  EU(action_{max}) - EU(action_n) > \delta_{action_{max}} + \delta_{action_{n}}
	\end{equation}

\noindent
Thus, to compare Process A and Process B it is necessary to consider all terms that are different. Therefore, the error term $\delta_a$ for Process A is related to directly sampling values for the
expected utilities, while in Process B the expected	utility is determined indirectly. For this reason, in Process B it is necessary to apply error propagation rules:

 \begin{equation} \label{eq:comp11}
	 EU(a|e)+\delta_{EU(a|e)}= \sum_{R} (P(Result=r|a,e)+\delta_b)*U(r) 
\end{equation}
	
\noindent Before applying error propagation to this equation, we need to normalize it so that $EU(a|e)/k$ is equal to $P(a)$.

 \begin{equation} \label{eq:comp12}
	 P(a)+\delta_a= \sum_{R} (P(Result=r|a,e)+\delta_b)*F(r) 
\end{equation}

\noindent where the normalization function ($F(r)$) is expressed as,

\begin{equation} \label{eq:comp13}
	 F(r)=\frac{U(r)}{\sum_a\sum_r U(r)*P(r|a,e)}
\end{equation}

\noindent Here, again, the mean value of $U(r)$ is used:

\begin{equation} \label{eq:comp14}
	 F(r)=\frac{U(r)}{\sum_a\sum_r P(r|a,e)*U(r)}=\frac{U(r)}{\sum_a \frac{1}{N_r}} =\frac{U(r)}{\frac{N_a}{N_r}}= \frac{N_r*U(r)}{N_a}
\end{equation}

\begin{equation} \label{eq:compl14}
	 F(r)= \frac{1}{N_a}
\end{equation}

\noindent Expressing the equation that determines the error term $\delta_a$ as a function of the error term $\delta_b$ yields

 \begin{equation} \label{eq:comp15}
	 \delta_a=\sqrt{ \sum_{R} \delta_b^2*F(r)^2}
\end{equation}

\noindent Using equation \eqref{eq:compl14} we obtain:

\begin{equation} \label{eq:comp17}
	 \delta_a=\sqrt{ \sum_{R} \delta_b^2*{(\frac{1}{N_a})}^2}
\end{equation}

\noindent Then, assuming that $\delta_b$ is similar, which is in favor of Process B because it minimizes the $\delta_a$ term:

\begin{equation} \label{eq:comp18}
	 \delta_a=\sqrt{ N_r*\delta_b^2*{(\frac{1}{N_a})}^2}
\end{equation}

\noindent yielding,

\begin{equation} \label{eq:comp19}
	 \delta_a={(\frac{\sqrt{ N_r}}{N_a})}*\delta_b
\end{equation}
	
\noindent With the relation between the error terms determined, it is possible to compare the difference of the computational effort involved in both processes, assuming again the mean terms for the probabilities:

\begin{equation}
    \sqrt{\frac{N_a}{N_r}}*\frac{A_{r,\alpha}*\frac{1}{N_r}*(1-\frac{1}{N_r})*\delta_a^2*N_a}{A_{a,\alpha}*\frac{1}{N_a}*(1-\frac{1}{N_a})*\delta_b^2} + \dfrac{2*N_a*N_r}{n*2^m*\sqrt{\dfrac{N_r}{P(e)}}*\dfrac{A_{a,\alpha}*\frac{1}{N_a}*(1-\frac{1}{N_a})}{\delta_a^2}}
\end{equation}

\noindent  Let us call the term on the right,

\begin{equation}
    t_1= \dfrac{2*N_a*N_r}{n*2^m*\sqrt{\dfrac{N_r}{P(e)}}*\dfrac{A_{a,\alpha}*\frac{1}{N_a}*(1-\frac{1}{N_a})}{\delta_a^2}}
\end{equation}

\noindent Using \ref{eq:comp19},

\begin{equation}
    \sqrt{\frac{N_a}{N_r}}*\frac{N_r}{N_a}*\frac{A_{r,\alpha}*\frac{1}{N_r}*(1-\frac{1}{N_r})}{A_{a,\alpha}*\frac{1}{N_a}*(1-\frac{1}{N_a})} + t_1
\end{equation}

\noindent also,

\begin{equation}
     \sqrt{\frac{N_r}{N_a}}*\frac{A_{r,\alpha}*\frac{1}{N_r}*(1-\frac{1}{N_r})}{A_{a,\alpha}*\frac{1}{N_a}*(1-\frac{1}{N_a})}+ t_1
\end{equation}

\noindent and,

\begin{equation}
     \sqrt{\frac{N_r}{N_a}}*\frac{A_{r,\alpha}*(\frac{1}{N_r}-\frac{1}{N_r^2})}{A_{a,\alpha}*(\frac{1}{N_a}-\frac{1}{N_a^2})} + t_1
\end{equation}

From \cite{Inglot2010} we obtain a lower bound for $A_{\alpha,k} $. Althouhg these terms are different for distinct values of $\alpha$, we consider the one where $\alpha$ is not leaning to zero too fast. Thus,

\begin{equation}
    A_{\alpha,k} \geq k + 2*\log{\frac{1}{\alpha}} - \frac{5}{2}
\end{equation}

\noindent With this equation it is possible to define a better value for the difference between the computational efforts,

\begin{equation}
\sqrt{\frac{N_r}{N_a}}*\frac{(N_r + 2*\log{\frac{1}{\alpha}} - \frac{7}{2})*(\frac{1}{N_r}-\frac{1}{N_r^2})}{(N_a + 2*\log{\frac{1}{\alpha}}-\frac{7}{2})*(\frac{1}{N_a}-\frac{1}{N_a^2})} + t_1
\end{equation}

\noindent As
\begin{equation}
 \lim_{N_r\to\infty} (N_r + 2*\log{\frac{1}{\alpha}} - \frac{7}{2})*(\frac{1}{N_r}-\frac{1}{N_r^2})=1 
\end{equation} 
 
\noindent and, 

\begin{equation} \label{eq:eq35}
 \lim_{N_a\to\infty} (N_a + 2*\log{\frac{1}{\alpha}}-\frac{7}{2})*(\frac{1}{N_a}-\frac{1}{N_a^2})=1
\end{equation}

\noindent it is possible to approximate the expression to
\begin{equation}
\sqrt{\frac{N_r}{N_a}}*\frac{(N_r + 2*\log{\frac{1}{\alpha}} - \frac{7}{2})*(\frac{1}{N_r}-\frac{1}{N_r^2})}{(N_a + 2*\log{\frac{1}{\alpha}}-\frac{7}{2})*(\frac{1}{N_a}-\frac{1}{N_a^2})} + t_1 \approx \sqrt{\frac{N_r}{N_a}} +\dfrac{2*N_a*\sqrt{N_r*P(e)}*\delta_a^2}{n*2^m}
\end{equation}

\noindent Writing the term of $\delta_a$ as a function of its dimension and a factor that adjusts the precision,

\begin{equation}
    \delta_a = \dfrac{1}{c*N_a}
\end{equation}

\noindent we obtain,

\begin{equation}
 \sqrt{\frac{N_r}{N_a}} +\dfrac{2*\sqrt{N_r*P(e)}}{N_a*c^2*n*2^m}
\end{equation}

\noindent Rewriting this the expression as,

\begin{equation}
 \sqrt{\frac{N_r}{N_a}} * (1+\dfrac{2*\sqrt{P(e)}}{\sqrt{N_a}*c^2*n*2^m})
\end{equation}

\noindent Because,

\begin{equation}
   \dfrac{2*\sqrt{P(e)}}{\sqrt{N_a}*c^2*n*2^m} \geq 0
\end{equation}

\noindent for any value of the composing variables, then,

\begin{equation}
 \sqrt{\frac{N_r}{N_a}} * (1+\dfrac{2*\sqrt{P(e)}}{\sqrt{N_a}*c^2*n*2^m}) \geq \sqrt{\frac{N_r}{N_a}}
\end{equation}

\noindent we prove that the relation between Process A and B is under bounded by,

\begin{equation}
    \sqrt{\frac{N_r}{N_a}}
\end{equation}

\end{document}